\documentclass[journal]{IEEEtran}

\usepackage{enumerate}
\usepackage[normalem]{ulem}
\usepackage{graphicx}
\usepackage{url}
\usepackage{amsthm}
\usepackage{comment}
\usepackage{balance}
\graphicspath{ {./images/} }
\theoremstyle{definition}

\begin{document}
\title{Developing Cyber Peacekeeping: Observation, Monitoring and Reporting}

\author{Michael~Robinson, Kevin~Jones, Helge~Janicke and  		Leandros~Maglaras, Senior Member, IEEE
\thanks{M. Robinson is with Airbus, Newport, United Kingdom (michael.mi.robinson@airbus.com)}% <-this % stops a space
\thanks{K. Jones is with Airbus, Newport, United Kingdom (kevin.jones@airbus.com)}%
\thanks{H. Janicke is with the Software Technology Research Laboratory, De Montfort University, Leicester, United Kingdom (heljanic@dmu.ac.uk)}%
\thanks{L. Maglaras is with the Software Technology Research Laboratory, De Montfort University, Leicester, United Kingdom (leandrosmag@gmail.com)}}%

% The paper headers
%\markboth{May~2018}%
%{Shell \MakeLowercase{\textit{et al.}}: Designing Cyber Peacekeeping}

% make the title area
\maketitle

\begin{abstract}
Cyber peacekeeping is an emerging and multi-disciplinary field of research, touching upon technical, political and societal domains of thought.  In this article we build upon previous works by developing the cyber peacekeeping activity of observation, monitoring and reporting.  We take a practical approach: describing a scenario in which two countries request UN support in drawing up and overseeing a ceasefire which includes cyber terms.  We explore how a cyber peacekeeping operation could start up and discuss the challenges it will face.  The article makes a number of proposals, including the use of a virtual collaborative environment to bring multiple benefits.  We conclude by summarising our findings, and describing where further work lies.
\end{abstract}


% For peer review papers, you can put extra information on the cover
% page as needed:
% \ifCLASSOPTIONpeerreview
% \begin{center} \bfseries EDICS Category: 3-BBND \end{center}
% \fi
%
% For peerreview papers, this IEEEtran command inserts a page break and
% creates the second title. It will be ignored for other modes.
\IEEEpeerreviewmaketitle

\section{Introduction}
\IEEEPARstart{T}{h}e United Nations conducts peace operations around the world, aiming to maintain peace and security in conflict torn areas.  Whilst early operations were largely successful, the changing nature of warfare and conflict has often left the UN struggling to adapt.  In this article, we make a contribution towards efforts to plan for the next evolution in conflict:  cyber warfare.  It is now widely accepted that cyber warfare will be a component of future conflicts, and much research has been devoted to its study.  Despite the vast amount of research relating to cyber warfare, there has been relatively little on its impact towards peace operations.

Previous work on this topic has sought to define the concept of cyber peacekeeping~\cite{Cahill2003,Akatyev2015,Robinson2018}.  The goal of this paper is to follow up specifically on work by Robinson et al.~\cite{Robinson2018}.  We build upon this work by exploring the practicalities of starting up a cyber peacekeeping component and conducting the initial task of observation, monitoring and reporting.

\section{Methodology}\label{Methodology}
In Robinson et al.~\cite{Robinson2018}, each United Nations (UN) peacekeeping activity was briefly examined for feasibility and value in a cyber warfare context. The aim was to cover the breadth of peacekeeping, and to pick out specific activities which would be both valuable and feasible to perform.  This article builds upon this work, by taking the activity of cyber observation, monitoring and reporting (OMR) and discussing how it could be carried out at a practical level.  To do this, we also describe the mission start-up process which leads to cyber OMR.

To guide our discussion, we begin by describing a fictional conflict scenario where cyber warfare has been used and is of concern to the involved parties.  This scenario is described in section\ref{scenario}.

\section{Scenario}\label{scenario}
The neighbouring states of Country A and Country B have a history of conflict, which has traditionally been confined to the domains of air, land and sea.  However, in recent months the cyber domain has also been used in a warfare capacity.  The cyber warfare aspect of the conflict has included both hard and soft attacks.  By "hard" we mean attacks such as denial of service and the planting of malware into sites such as critical infrastructure (power grid, water supply, transport systems etc.).  This cyber warfare has been particularly damaging to both sides, interrupting the provision of basic services.  Power outages have been common in both nations, and public trust in the water supply is damaged after high profile cases of improper water treatment.  "Soft" attacks have included the spreading of misinformation to the public via cyber means and accusations of electoral interference.  This combination of hard and soft cyber attacks, has placed both countries on the verge of collapse as a lack of basic services combined with suspicion over the legitimacy of election results has led to civil unrest.

Both countries seek an end to the situation and express a wish for a halt to the conflict.  They request UN assistance in drawing up, implementing and verifying compliance with a ceasefire agreement.   UN peacemakers begin work to draw up traditional terms such as withdrawal of armed forces to a defined boundary, the holding of negotiations to reach a long term sustainable peace and a process of disarmament by both parties.  However, both parties have been significantly damaged by the cyber warfare aspects of the conflict and agree that the ceasefire agreement should also contain cyber terms to end it.  Both sides look to the UN peacemakers for leadership and advice in this domain.

\subsection{Cyber Peacemaking}\label{peacemaking}
In our scenario, both parties look to the UN for assistance in drawing up cyber related ceasefire terms and overseeing their implementation.  This will be a new area for the UN, and it is therefore necessary to briefly discuss cyber peacemaking and the kinds of terms which may be agreed upon.  The process of mediating and drawing up an effective and lasting ceasefire agreement is a complex and challenging task~\cite{Akebo2016,Johnson2012}.  A full discussion of this complexity is beyond the scope of this article.  Instead, we focus upon the possibilities of cyber related ceasefire terms.  This in itself will become a field of research and practice in the future; what we provide here is only the foundation.

A clear starting point for thinking about cyber related ceasefire terms is to propose that both parties cease launching cyber attacks.  This aligns with one of the three core ceasefire goals:  cessation of hostilities~\cite{ceasefirebook2013}.  In practice such a term would be difficult to monitor due to the problem of attribution in cyberspace~\cite{cook2016attribution}.  As discussed in Robinson et al.~\cite{Robinson2018} and in many other works~\cite{Vihul2014,Berghel2017}, this is the technical challenge of gathering unequivocal proof that cyber attack X came from party Y.  In short, it is difficult to present proof that would pass even the minimal of evidentiary standards~\cite{Berghel2017}.  This is not particularly a problem for nations; they are free to attribute attacks without fear of having to reveal their sources or open their evidence to independent inspection.  This leads to what has been named faith-based attribution: nonscientific analysis that leads to untestable attribution~\cite{Berghel2017}.  UN peacekeeping cannot and should not employ such methods.  It is an activity based upon the trust of both parties and transparency.  It is therefore essential that claims of ceasefire violations must be backed up by open evidence.  It has consequently been argued that peacemakers should simply avoid any cyber terms which require solid, verifiable attribution~\cite{Robinson2018}.

This conclusion limits the types of terms which could be included in a cyber ceasefire agreement.  For the purposes of our scenario, we can take the examples given by Robinson et al.~\cite{Robinson2018}, and add one more:

\begin{enumerate}
	\item Both countries to provide full assistance to UN cyber peacekeepers in dismantling botnets and other sources of denial of service attacks which are physically located in their borders.
	\item Both countries to cooperate with the UN in the prevention of cybercrime/spoiler attacks which are originating from within their  borders.
	\item Declaration of information stolen during the conflict.
	\item Declaration of systems compromised and assistance with returning control to rightful owners.  Caution must be used here because there may be a dispute about who the rightful owners of certain systems are.
	\item Declaration of known vulnerabilities in critical infrastructure.
	\item Remote disabling of malware (if possible) or assistance in locating and removing malware.
    \item The right for each party to request UN cyber peacekeeper monitoring of particular sites.
\end{enumerate}

These ceasefire terms avoid the attribution problem because they all involve person to person cooperation that can be observed and monitored.  Peacekeepers can observe how cooperative each party is in helping to dismantle botnets, disable malware, return control of systems and declare vulnerabilities.  Regardless of the result of such cooperation, clear demonstrations of transparency, honesty and openness in assisting cyber peacekeepers would be seen as fulfilment of the agreement.  Conversely, inaction, opaqueness or refusal to cooperate can be indicators of a violation.  Measuring the levels of compliance in this way will be much more productive to peace and a better use of resources in comparison to engaging into debates about proof with each party.  It must be noted that choosing to sidestep the challenge of attribution does not mean cyber peacekeepers will simply ignore the problem of cyber attacks.  As will be discussed, they will be handled in a defensive manner.

It is noted that these are a simplification of ceasefire terms.  As noted in ceasefire literature~\cite{Haysom2004,ceasefirebook2013} there is no room for ambiguity, and every term such as denial of service, botnet and cyber-crime must be clearly defined and agreed with both parties.  The terms presented here are therefore just examples, chosen for the purposes of describing how cyber OMR could be performed.

\section{Mission Start-Up}
With a ceasefire in place, the UN Security Council is in a position to authorise a UN peacekeeping operation which contains a cyber component.  We therefore explore the practicalities of starting up the cyber component of such a mission.  Whilst UN processes are complex, we propose that a cyber component can fit into existing processes without the need for any significant changes to the established ways of working.  This is because the processes themselves have been designed with the aim of unifying multiple disparate entities.  We begin with a discussion on securing the necessary cyber expertise.

\subsection{Finding Cyber Expertise}\label{findingexpertise}
Our scenario represents a situation the UN may find itself in:  being asked to guide the drawing up of a ceasefire agreement which meaningfully addresses cyber warfare and overseeing its implementation.  If unprepared, the UN may struggle to secure the expertise at short notice, leading to delays in the peace process.  This is undesirable as once started, a peace process must progress quickly to capitalise upon the honeymoon period~\cite{UN2003}.  Similarly, securing the wrong types of expertise may lead to failed implementation and a relapse into conflict.  It is therefore essential to think about how the UN could secure suitable cyber expertise ahead of time.  To assist with this task, we first describe how the UN secures expertise presently.

The UN has no standing army or police force, and must request contributions of troops, police and observers from UN member states who act as Troop Contributing Countries (TCCs) and Police Contributing Countries (PCCs)~\cite{UN2008}.  As of February 2018, the UN had just over 100,000 contributed police, military experts and troops from 123 nations~\cite{UNdata2018}.  Although the word contribution suggests that they are charitably donated, nations contributing troops are reimbursed at a standard rate, which in 2018 is US\$1,332 per soldier per month~\cite{UNfunded2018}.  Numerous works exist which explore why states do or do not choose to contribute personnel~\cite{Bellamy2013,zhengyu2011,bove2011}.

The UN also maintains a pool of civilian staff.  As of 2018 the UN had more than 15000 civilians serving in peacekeeping operations around the world~\cite{UNcivilians2018}.  These civilians fulfil many roles not suited to either troops or police, such as providing general administrative assistance, acting as public information officers or providing specific expertise such as logistics or ICT knowledge.  Civilians can serve as international staff, national staff from the host country, as UN volunteers, consultants or contractors~\cite{UNcivilians2018}.  

Clearly the UN already has the systems in place to acquire expertise from multiple sources:  it can tap into national militaries and police forces around the world or from civilian sources where necessary.  When considering cyber peacekeeping, we propose that cyber expertise could come from all of these sources.  States can be a source of cyber peacekeepers by contributing military cyber warfare troops and cyber crime police officers, but expertise can also be found in non-governmental organisations, private industry, charities and volunteers.  It is therefore possible to define multiple sources of cyber peacekeepers:

\begin{itemize}
	\item Cyber Contributing Countries (CCCs) - States which contribute uniformed cyber peacekeepers from the police or military.
	\item Cyber Contributing Organisations (CCOs) - Organisations which contribute civilian cyber peacekeepers from their workforce.
    \item Volunteers - People with cyber expertise who volunteer their time.
    \item Full time UN cyber staff - civilians recruited as employees by the UN. 
\end{itemize}

Aside from these traditional approaches to securing personnel, there is ongoing debate regarding the use of private security companies (PSCs) in UN peacekeeping operations.  The use of private security companies in conflict areas is not new, with companies such as Blackwater providing armed security in Iraq~\cite{Elsea2008}.  To date, the UN has avoided using PSCs for front line activities, but has contracted companies such as ArmorGroup for mine action duties with successful results~\cite{Bellamy2009}.  With a willingness to use PSCs, their potential as a reliable source of cyber peacekeepers is clear but caution must be used since the disadvantages of PSCs are well documented~\cite{Bellamy2009}.

An advantage the UN has in searching for cyber expertise is that cyber peacekeepers will be required in much lower numbers than troops or police.  While a border may require thousands of troops to guard and patrol it, a network can be monitored by a handful of cyber peacekeepers, utilising technology to automate a number of monitoring tasks.  This lower number of required staff will slightly mitigate the many problems described next.

Although there are many potential sources of cyber peacekeepers in theory, they will likely be challenging to secure in practice.  The reasons for this are political, societal and economic in nature.  To begin, we briefly discuss the existing challenges of securing troop and police contributions.

The problems faced by the UN in securing good quality contributions is well documented~\cite{Serafino2004,Durch2003}.  Nations with highly trained troops and police are unwilling to face a shortage at home by sending them abroad to potentially dangerous regions~\cite{Bellamy2013}.  Those nations with inexpensive personnel are more likely to contribute, but the incentive here is often economic:  the compensation provided by the UN per soldier or officer is often more than their cost~\cite{Bellamy2013}.  For example, some of the most significant contributions have come from nations such as Bangladesh, Pakistan, India, Nigeria and Ghana~\cite{Bellamy2013}.  The US, UK, France, China and Russia all have a record of low personnel contributions to UN peacekeeping missions~\cite{Bellamy2013}.  Whilst economics plays a part in this discrepancy, it is not the only factor.  The decision on whether to contribute can depend on many factors such as if it is in their own national interest (e.g. the instability is on their border and could spillover) and the level of toleration for casualties (the US withdrawal from Somalia in 1994 was partially due to a low tolerance for the loss of US troops)~\cite{bove2011}.

When looking to the cyber domain, we must navigate these factors to evaluate how likely contributions of cyber personnel will be.

From an economic perspective, the indicators are not good.  As the threat of cyber warfare increases, nations are competing to secure good quality cyber expertise in a market with limited supply.  In 2015 it was reported that the US Cyber Command only has half the staff required, citing competition with the private sector as an obstacle towards securing the staff numbers it needs~\cite{Sternstein2015}.  With nations and businesses around the world competing for a limited supply of good quality cyber personnel and incidences of cyber warfare and crime increasing, cyber peacekeepers will be an expensive asset.

Looking from a political angle, the willingness to suffer a cyber skills shortage at home for the benefit of UN peacekeeping abroad is likely to be limited.  Even if a highly cyber developed nation could technically spare cyber personnel, it is debatable whether they would allow their staff to work alongside staff from another cyber competing nation.  This is significant in a political environment where tensions over cyber warfare currently run high between some of the most cyber developed nations~\cite{Libicki2016,Shaikh2018}.  Nations may prefer keep their cyber personnel at home, or to limit sharing with trusted partners such as NATO or the African Union.

Although there are clear challenges, some aspects may encourage contributions.  For example, proximity to the conflict area may be a driver of cyber peacekeeper contribution:  critical infrastructure at risk of failing could potentially impact neighbouring states.  In this regard, there is a self interest to contribute the expertise.  Furthermore, it has been noted in the literature that peacekeepers often stand to benefit professionally from the experience~\cite{Thakur2007}.  Peacekeepers also largely report their duties to be satisfying~\cite{Heinecken2012} and studies have shown that the experience of peacekeeping has a beneficial effect on quality of life~\cite{loscalzo2018}.  With this in mind, cyber peacekeeping may be attractive to civilians who are looking for both job satisfaction and a level of professional development that otherwise might not be open to them.

Although the UN will be aiming to secure relatively small numbers of cyber peacekeepers, the issues described will be a significant challenge.  Nations may be unwilling to contribute personnel and it remains to be seen how much civilian interest will emerge.  This challenge cannot go ignored, and methods to encourage contributions and civilian recruitment must be built into the design.

\subsection{Fitting cyber into the organisation}\label{organisation}
While securing cyber talent will be challenging, we assume that the UN will be able to secure some.  With this assumption, we must next consider how these cyber staff will be organised, and where they will be placed into the existing organisational structure.  Figure~\ref{fig:CommandStructureTraditional} shows a simplified version of the existing structure:

\begin{figure}[ht]
\centering
\includegraphics[width=\columnwidth]{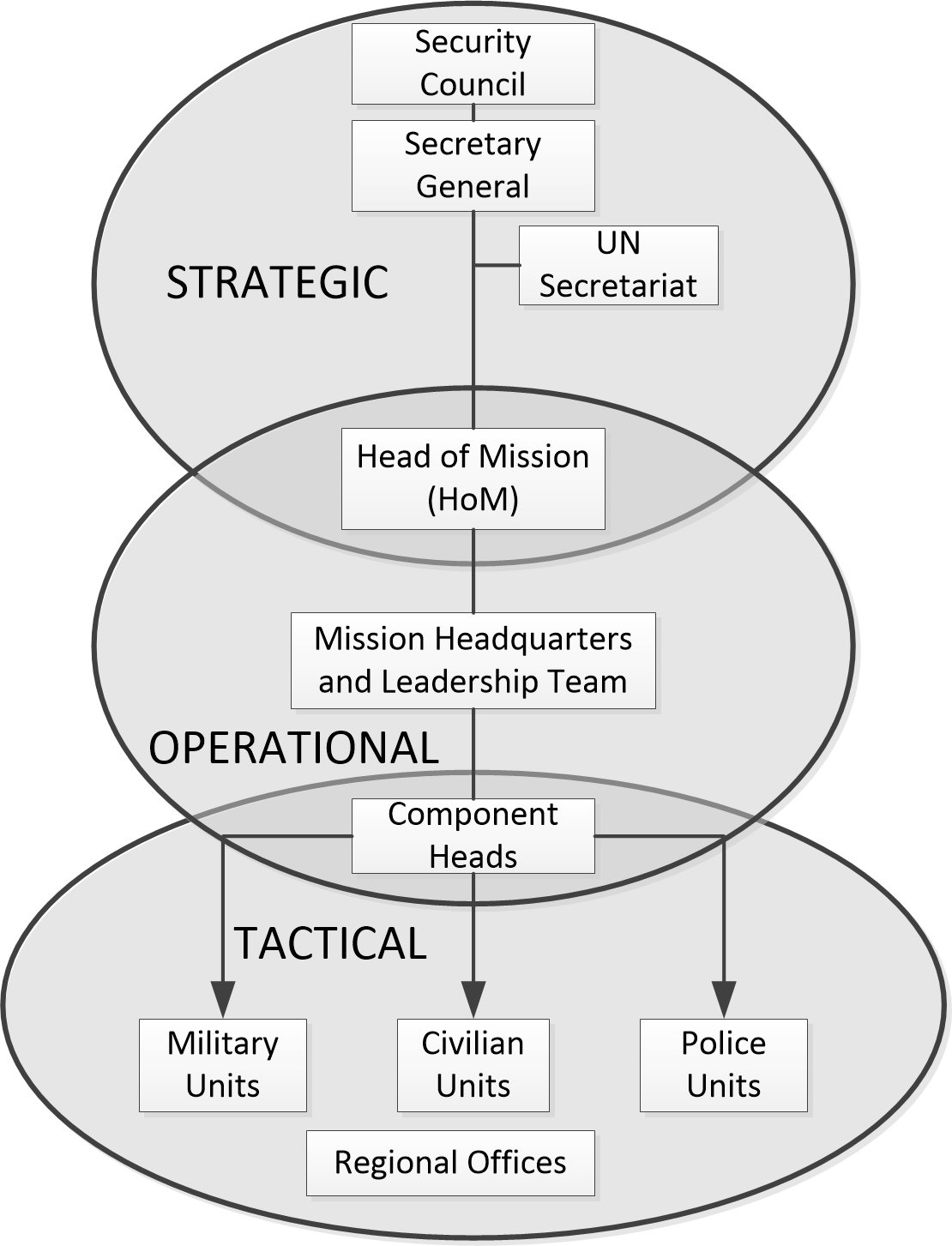}
\caption{Traditional UN Peacekeeping Organisational Structure}
\label{fig:CommandStructureTraditional}
\end{figure}

At the tactical level there are currently military, civilian and police components.  Each of these units is overseen by a component head who reports to a leadership team at the mission headquarters.  At the tactical level, an immediate question becomes apparent: should cyber peacekeepers be considered as military, police or civilian units?

\begin{figure}[ht]
\centering
\includegraphics[width=\columnwidth]{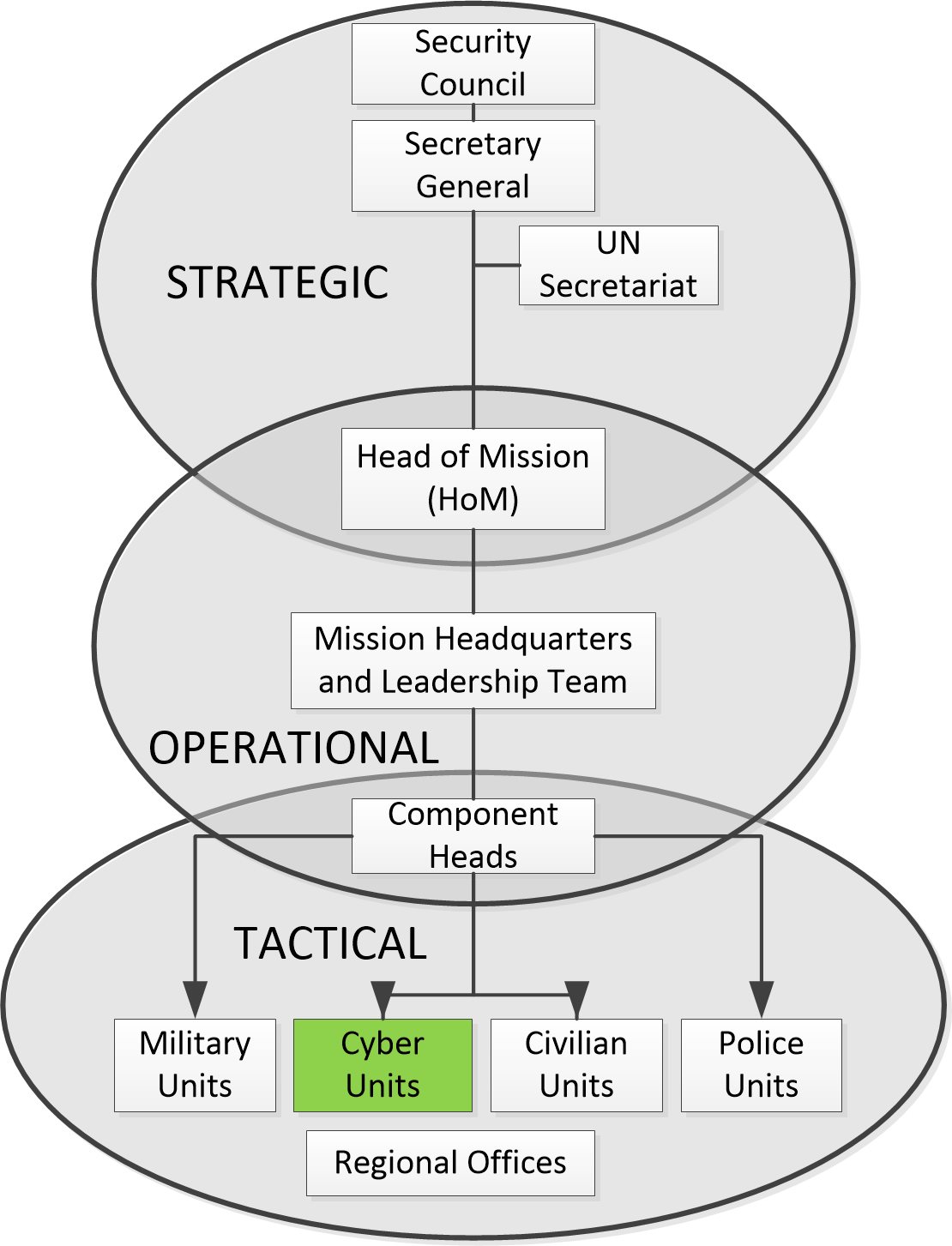}
\caption{Proposed UN Peacekeeping Organisational Structure (Cyber Units Added)}
\label{fig:CommandStructureNew}
\end{figure}

Based on our discussion in section~\ref{findingexpertise}, we have already proposed that cyber peacekeepers will come from multiple sources:  some will be from the military, some from police forces and some from civilian backgrounds.  One approach would be to simply assign cyber peacekeepers to the unit which matches their background.  For example, a cyber peacekeeper from a national military would be assigned to a military unit whilst civilians would be assigned to a civilian unit.  This approach does have its advantages (for example, a military cyber peacekeeper could potentially be armed and supported by infantry) but it would also potentially damage communication and coordination between cyber peacekeepers.  We therefore propose that due to the unique nature of cyber peacekeepers, a new type of unit is created: cyber units.  Placing cyber peacekeepers into a cyber unit would facilitate cohesion, regardless of their source.  It would also open up a new role, the head of cyber component, to oversee and be responsible for the tactical cyber aspects of a particular operation.  Figure~\ref{fig:CommandStructureNew} shows where a cyber component would fit into the organisational structure.

Whilst the head of cyber component will report to the leadership team and head of mission for operational planning and reporting, it is envisioned that cyber experts will also be present at the strategic level to guide decision making in strategic aspects relating to cyber.

\subsection{Mission Planning}
\begin{figure}[ht]
\centering
\includegraphics[scale=1]{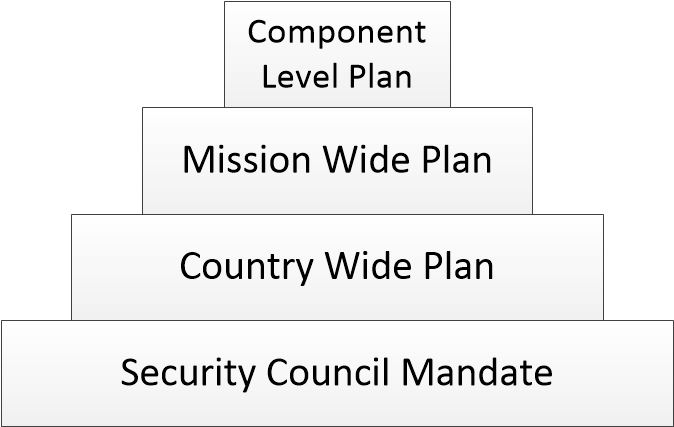}
\caption{UN Peacekeeping Plan Levels}
\label{fig:planlevels}
\end{figure}

The overall planning process for UN peacekeeping operations is complex, consisting of multiple plans and processes\cite{IAP2013,boutellis2013}.  A simplified overview of the plans that need to be developed and their hierarchy in relation to each other is shown in figure~\ref{fig:planlevels}.
  
Laying the foundation for other plans, the Security Council mandate sets out an overall political goal.  In our scenario, this would be to secure peace between the two countries.  This mandate is then used to formulate a country wide plan: for example, what is required in Country A to fulfil the mandate.  Mission wide plans then define how these political intents can be reached at the strategic level, giving statements on the type of activities that will be required.  Finally, the component level plans use the mission wide plans to formulate plans at the tactical level:  specific actions, time lines, expected outcomes and so on.  In the next section, we explore how such a component plan could look for a cyber unit in Country A.

\section{The Cyber Component Plan}
In our scenario, the cyber component plan would set out the high level objective of the cyber component: to support the Security Council mandate by providing the cyber capability that is necessary to support the restoration of peace.  In particular, the priority of the cyber peacekeeping unit (CPU) will be to ensure compliance with the cyber terms in the ceasefire agreement.  This high level objective will then be broken down into smaller tasks, such as conducting technical assessment missions on proposed sites, forming links with local stakeholders and building an observation, monitoring and reporting capability at sites where the value towards maintaining peace is high.

\subsection{Technical Assessments}
One of the terms in our ceasefire agreement is the right for each party to request UN cyber monitoring of particular sites.  Country A requests monitoring of:

\begin{itemize}
	\item The president's office.
	\item The power grid.
    \item The control system for their flood defences.
\end{itemize}

These requests are initially made to UN peacemakers, but are passed to the head of cyber component for expert evaluation.  The head of cyber component considers the requests and launches technical assessment missions to evaluate which to grant.  Such assessments are not new to UN peacekeeping.  They are already well established in peacekeeping as a means to evaluate the feasibility and value of carrying out a particular task before it is agreed to.  The UN planning toolkit~\cite{UNPlanningToolkit2014} provides guidance on how such assessments should be conducted.  From a cyber component perspective, these assessment missions should evaluate:

\begin{enumerate}
	\item The specific ways in which monitoring this site would contribute towards peace (e.g. a cyber attack upon the site could threaten civilian lives or lead to state collapse).
    \item The level of local support: do local staff welcome the cyber peacekeepers and cooperate?  Are network diagrams made available for inspection and are staff forthcoming with assistance?
    \item Capacity to act: can the site be effectively monitored given the current funding, availability of cyber expertise and equipment?
\end{enumerate}

In practice this assessment would likely find limited value in monitoring the president's office and this request would be denied; however, for the purposes of describing how to conduct cyber OMR in its simplest form, it is a useful example.  This is because the president's office is a traditional ICT environment characterised by desktop computers, mobile devices, printers, databases and so on.  It is this kind of environment in which the majority of cyber security literature and understanding is based.   We will therefore assume that the request was granted and describe how a cyber OMR capability would be established in an ICT environment such as the president's office.

\section{Cyber OMR in ICT environments}
From the conclusions made in Robinson et al.~\cite{Robinson2018}, combined with our scenario's ceasefire terms, it is possible to be concise about what cyber OMR is trying to achieve in our scenario:

\begin{itemize}
    \item Detecting actions which violate the ceasefire agreement - our ceasefire terms were chosen to avoid attribution and to be monitorable through social interaction.
    \item Detecting violations of human rights - In our scenario this is primarily protecting civilians from harm (e.g. the right to life).
    \item Detecting changes in network structure and network traffic - by itself this does not bring direct value, but supports the other goals by raising situational awareness at a network.
\end{itemize}

The first goal does not require any technical discussion: peacekeepers already observe levels of cooperation and compliance and this happens on an interpersonal level.  Where more discussion is required is in regards to the second and third goals and the necessity to establish a technical monitoring capability at agreed sites.

\subsection{Establishing a monitoring capability}
Just as military units will look to establish an effective observational capability through the use of patrols, observation posts and checkpoints~\cite{UNIBAM1}, cyber units will look to establish an effective observational capability through cyber means.

The monitoring of computer networks is a well established domain of cyber security.  We can therefore look to existing literature and best practice for guidance on how we could establish this capability in a peacekeeping environment.  For example, we can turn to guidance from authors such as Bejtlich~\cite{Bejtlich2013} on how to establish suitable network security monitoring solutions in a traditional ICT environment.

This will initially involve a planning stage, where cyber peacekeepers consult with local staff and build a picture of the network, the information expected to be flowing in and out and any existing monitoring solutions.  In line with established doctrine, the emphasis in cyber OMR should be to engage local stakeholders such as IT staff and support them with the cyber knowledge, expertise and equipment they may lack, rather than taking over the site and dictating actions.  In our scenario, the UN is operating with consent of the host nation and all involved parties.  Additionally, technical assessment missions have already concluded that local staff are willing to assist and are ready to engage with the process.

If discussions lead to the conclusion that existing sensor coverage is insufficient, new sensors should be placed into the network to fill gaps in visibility.  These sensors can be network taps (dedicated devices added onto the line to intercept all traffic) or span ports at existing network devices (a port which replicates all data passing through a switch).

Once cyber peacekeepers and local staff are satisfied that monitoring coverage is sufficient, they can begin to monitor the network and report upon events which may impact the security of civilians.  The technical methods of interpreting the captured data is an established field of study, and cyber peacekeepers will be expected to exercise and share their expertise of network security monitoring here.  Numerous sources of guidance and best practice exist~\cite{Collins2014,Sanders2013,Sanders2017}, as well as various off-the-shelf software tools such as Security Information and Event Management (SIEM) products and packet sniffers.  Cyber peacekeepers will use their expertise to determine which tools will be best for the particular environment they are working in.

The key ingredients to success in this scenario are two-fold: the ability for cyber peacekeepers to build up rapport with local staff and for them to gain a familiarity with the network they are monitoring.  Where networks are small, this is not likely to be a problem.  Where a network is larger or highly geographically dispersed, these two goals will be harder to reach.  When considering systems such as the power grid, components can include power plants, transmission systems, distribution substations and more.  We must consider ways for a cyber peacekeeping unit to gain familiarity across such large sites both technically and socially.

We propose that the concept of areas of responsibility from existing peacekeeping documentation can be leveraged for this task.  Areas of Responsibility (AoRs) is a term used by UN infantry battalions to divide a geographical area into smaller areas for groups of infantry to patrol and observe~\cite{UNIBAM1,UNIBAM2}.  This makes it easier for commanders to assign troops efficiently and ensures that necessary areas are covered by the right amount of manpower.  It also allows the building of trust between peacekeepers and local people, due to the opportunity to build up familiarity over time and develop an appreciation of local issues.  It is proposed that the concept of AoRs would be suitable for cyber OMR, since it would allow cyber peacekeepers to focus on one area of the site.  The benefits of using AoRs are summarised as follows:

\begin{itemize}
  \item Splits the observational workload into manageable sub-areas.
	\item Allows a team to build up familiarisation with the network area they are monitoring (its structure, traffic patterns etc.)
	\item Enables a team to build rapport with local staff on that part of the network.
	\item Allows cyber peacekeepers with different areas expertise to focus on specific systems.
\end{itemize}

\begin{figure}[ht]
\centering
\includegraphics[width=\columnwidth]{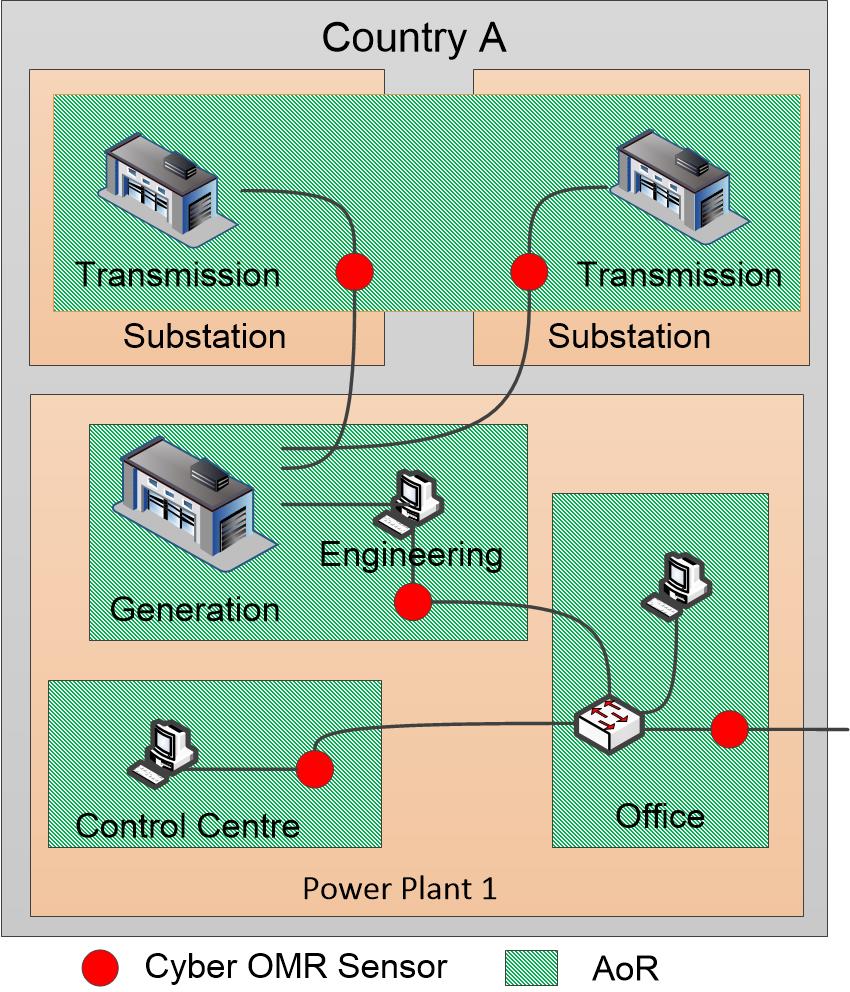}
\caption{Simple representation of areas of responsibility in cyber OMR}
\label{fig:AoRapproach}
\end{figure}

In the case of the power grid, a simplistic view of AoRs being leveraged is visualised in figure~\ref{fig:AoRapproach}.  In the figure, we see that multiple parts of the power grid have been split into observation areas.  Cyber peacekeepers with expertise in industrial control systems (ICS) can therefore be assigned to the generation or transmission AoRs, whilst those with more expertise in traditional ICT environments can be assigned to the office or control centre.  Once assigned, these staff will then be able to build up familiarity with both the systems they are monitoring and the local staff.

\subsection{Monitoring of an AoR}
With the establishment of AoRs at monitoring sites, cyber peacekeepers now have a technical capability to perform cyber OMR.  The next step is to consider how cyber peacekeepers use this capability to produce value for a peacekeeping operation.  It is proposed that there are two approaches towards the observation and monitoring of an AoR: local and remote. 

\subsubsection{Local}
The local approach involves cyber peacekeepers travelling to the AoR and performing their duties in person.  The advantage to this approach is that cyber peacekeepers will be able to build up a face to face relationship with local staff.  This facilitates trust between cyber peacekeepers and local stakeholders, which is regarded as vital to the success of any peacekeeping operation~\cite{UNIBAM2}.  Local cyber peacekeepers can also maintain the sensors as required.  The disadvantages of the local approach are numerous however.  Firstly, cyber peacekeepers would have to physically travel to the site; finding cyber peacekeepers who wish to relocate may be difficult.  At a personal level, the physical security of the location may be questionable, and work or family commitments may prevent it.  This may result in reduced civilian recruitment.  At an organisational level, the challenges regarding the global shortage of cyber personnel discussed in section~\ref{findingexpertise} may mean that governments and organisations are reluctant to be without cyber staff whilst they travel to a foreign country.  This issue has already been encountered by the UN in regards to securing police contributions~\cite{Serafino2004}.

It must be noted that the local method may be the only option available in certain cases.  For example, in cases where air-gapping theoretically exists, external connections may be prohibited and/or unavailable.  Similarly, the technical assessment missions and deployment of additional sensors if needed would further necessitate a local approach.

\subsubsection{Remote}
The second approach is the remote method, which capitalises upon an attribute of cyber warfare:  lack of geographical restriction~\cite{Robinson2014}.  Once a technical assessment mission has been completed locally and a monitoring capability established, it is envisioned that some monitoring could be performed remotely.  Sensors at the site can report back to a central server, whereby analysis of the collected data can be performed from any geographical location.

The advantages here are numerous and significant.  From the perspective of a contributing government or organisation, cyber experts they contribute are not being completely surrendered.  Contributors can agree to donate a limited portion of a cyber expert's time per day, the majority still being available to the contributor.  This is a significant benefit for the UN when trying to secure the necessary cyber talent in a highly competitive global market: if contributors are not losing access to the cyber expert, they will be more likely to contribute.  From the perspective of the cyber peacekeeper, the concern of physical security is also removed, and there is no need for the UN to fund the accommodation and living costs associated with the local approach.

The remote approach is not without disadvantages.  Even if the cyber peacekeeper is fully vetted and determined to be non-malicious, there is a risk when allowing cyber peacekeepers to perform their duties remotely with their own hardware.  For example, there is the potential for their system to be infected with malware which can lead to breaches of security.  To mitigate this issue, the UN may opt to send cyber peacekeepers a hardened, locked down and monitored system which is used purely for the purposes of cyber peacekeeping.  Similarly, the UN would be relying upon the public internet and peacekeeper's own internet providers to conduct cyber OMR.  At critical sites with the possibility of significant harm to civilians, this risk may be too high.  A possible mitigation of this risk would be to use a dedicated communications network.  Another disadvantage is that by being remote, there is potential to lose the face to face collaboration and rapport building that comes with the local approach.  To resolve this concern, it is proposed that a virtual collaborative environment (VCE) could be used.

\subsection{Virtual Collaborative Environment}
Virtual collaborative environments are digital spaces where remotely located people can come together and interact with each other and with virtual objects.  The benefits of VCEs are well established~\cite{Cherbakov2009,Redfern2002} and research into new applications for VCEs in areas such as science, education and business is ongoing~\cite{Zhang2003,Bouras2003}.

From a cyber peacekeeping perspective, potential off-the-shelf options include Vastpark, Protosphere, Second Life, Opensimulator and Open Wonderland.  Market reviews and research which describe the features and capabilities of these software options are available~\cite{Cherbakov2009,Surakka2011}.  A further option is for a custom solution to be developed using an engine such as Unity~\cite{unitywebsite}.  It is proposed that any VCE used by cyber peacekeeping must fulfil the following non-functional requirements:

\begin{enumerate}
	\item Scalable - The VCE should be suitable for both small and large scale cyber peacekeeping activities, and allow for changes in scale without disruption to the operation.
	\item Robust - The VCE must be reliable, accepting a high number of concurrent users with no failures of availability.  Considering the nature of cyber peacekeeping, it must also be secure from cyber attacks such as denial of service and man in the middle attacks.
	\item Secure - Sensitive information will be contained in the VCE. It must have access control, secure communications and auditing features.
\end{enumerate}

It must also fulfil the following functional requirements:

\begin{enumerate}
	\item Resource sharing - Cyber peacekeepers will need to examine sensor data together and study information from a variety of sources inside the VCE.
	\item VOIP - Allowing cyber peacekeepers to communicate with each other and local stakeholders in real-time.
	\item Reporting - The reporting system should be available from inside the VCE.
\end{enumerate}

In line with the goal of this paper to provide practical guidance, we developed a proof of concept VCE.  OpenSimulator was chosen for this since assets were readily available online and the set-up process was straight forward.  A basic world was built using terrain found at the OpenVCE website~\cite{openvce2015}.

\begin{figure}[ht]
\centering
\includegraphics[width=\columnwidth]{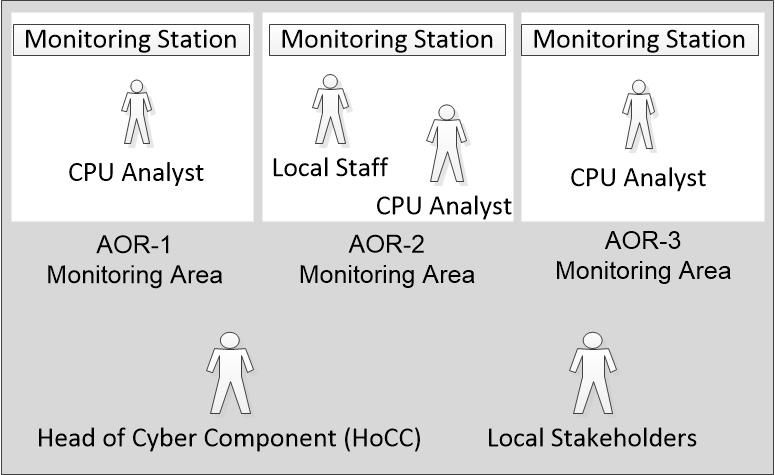}
\caption{VCE Layout Type 1}
\label{fig:VCE-Layout-Type1}
\end{figure}

Defining the requirements of a cyber peacekeeping VCE allowed the drawing up of potential layouts.  Figure~\ref{fig:VCE-Layout-Type1} shows the first design of a cyber peacekeeping VCE.  The Type 1 VCE is a simple layout whereby each AoR has a monitoring area, showing the sensor data and other information in real-time to cyber peacekeepers.  Cyber peacekeepers can log in to the system from anywhere in the world, gain access to the sensor data and communicate in real time with other cyber peacekeepers and stakeholders.  The Head of Cyber Component (HoCC) and his or her team is also present in the VCE if needed (for example, in times of crisis).  This infrastructure allows all parties to achieve situational awareness, communicate with each other in real time, cross correlate events and provide assistance or inspect incidents as they happen.  Local stakeholders are invited into the environment to communicate, contribute and witness how the monitoring is progressing.  This encourages transparency in the operation and fosters trust.  Liaisons from other unit types such as police and military can also be present, allowing for timely and unified coordination of actions across the whole operation.  Note that cyber peacekeepers are not strictly confined to their station, and may assist and communicate freely with other people as needed.

There are advantages and disadvantages to the Type 1 layout.  On the plus side, it is a simple design which fulfils the previous criteria.  All AoRs are monitored and staff can communicate and assist each other freely.  Arguably the most significant disadvantage is that it is resource intensive and fatiguing for staff.  Using this design, each AoR is manned 24/7 with cyber peacekeepers being required inside the VCE and performing real time monitoring.  This observation therefore raised the question of whether the monitoring needs to be real time or if it can be performed in batches (for example, reviewed every X hours).

\begin{figure}[ht]
\centering
\includegraphics[width=\columnwidth]{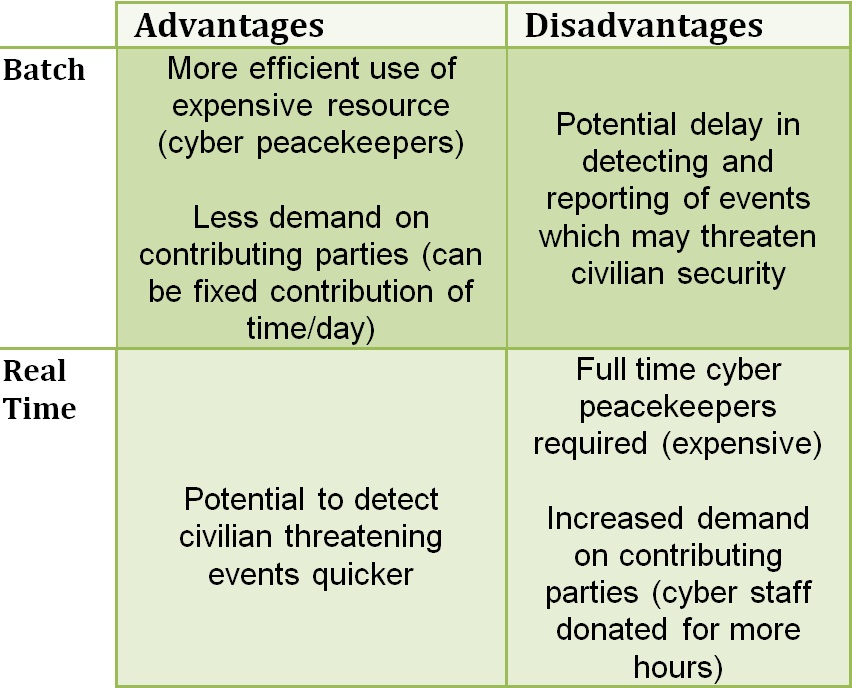}
\caption{Real-time vs. Batch Monitoring}
\label{fig:realtimevsbatchtable}
\end{figure}

To answer this question, we turn to existing cyber security literature.  Bejtlich~\cite{Bejtlich2004} states that the decision to use real-time or batch analysis during network security monitoring primarily depends upon management's expectations on timeliness of reports.  This is a reasonable stance, since in a private organisation it is management who balance up the risks versus costs and reach a conclusion that suits their organisation.  Peacekeeping is different; there are multiple stakeholders who each have their own expectations of what the operation should deliver.  Taking our scenario, Country A might be satisfied with daily reports containing summaries from the previous 24 hours.  Country B might have different expectations, and expect to be informed the moment a violation is detected.  Local variations may exist such as the the power grid demanding daily reports whilst the operator of the flood defences may desire them every four hours.  There are also additional considerations such as the potential for civilian harm if an attack is not observed quick enough.  In this regard, peacekeepers themselves may have their own view on how regularly data should be reviewed.  The pros and cons of real time and batch monitoring during cyber peacekeeping are shown in figure~\ref{fig:realtimevsbatchtable}.

The major advantage of real-time monitoring is that safety critical events can be detected quickly.  Although OMR is a passive activity in both kinetic and cyber peacekeeping, it has been repeatedly emphasised that impartiality does not mean neutrality~\cite{Donald2002}.  UN peacekeepers have a duty to intervene if they witness and are able to prevent a threat to the security of civilians~\cite{UN2008}.  This is a compelling justification for cyber OMR being real time, particularly on safety critical networks such as air traffic control, dams, and nuclear facilities.  This must be balanced against the personnel issues raised in section~\ref{findingexpertise}: cyber expertise will be expensive to secure, civilians will be difficult to recruit and contributing nations may be reluctant to contribute staff if they are required for extensive periods of time.

\begin{figure}[ht]
\centering
\includegraphics[width=\columnwidth]{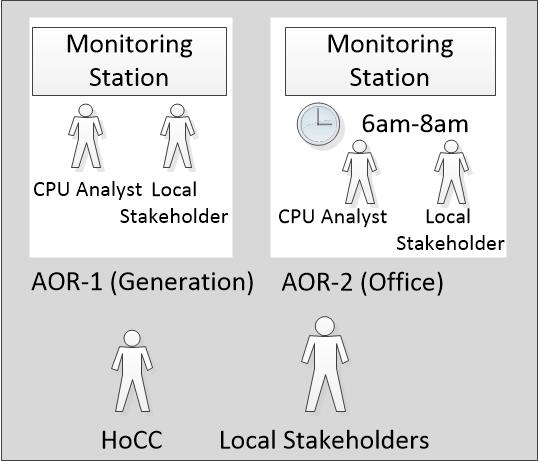}
\caption{VCE Layout Type 2}
\label{fig:VCE-Layout-Type2}
\end{figure}

Looking at the arguments for and against, it is concluded that cyber OMR should use both real-time and batch monitoring.  Batch monitoring should be the default choice, but technical assessment missions can recommend real-time if it is necessary to prevent harm to civilians or the wider peace process (e.g. potential for state collapse or destabilisation).  Relating this back to our scenario, the technical assessment mission at the power grid has concluded that whilst the generation AoRs will require real-time monitoring, the office would be suitable for batch monitoring every 24 hours.  Figure~\ref{fig:VCE-Layout-Type2} shows how this could be organised in the VCE.

Here cyber peacekeepers are providing real-time coverage to AoR-1, along with others who log in every 24 hours and perform batch monitoring in pairs for a period of two hours.  This pairing of staff allows them to discuss alerts and events between themselves before raising a report up the command chain.  It is less resource intensive, with the cyber peacekeepers and local staff only being required for the time it takes to work through the events from the previous 24 hours.

The significant benefit of this approach is the minimisation of the time required from cyber peacekeepers.  By using batch monitoring where possible the UN is not only seeking to minimise the number of cyber peacekeepers required, but also the amount of time they are required for.  The time commitment of two hours per day is only an example, and in reality will vary depending upon the complexity of the AoR.  The attraction of this limited time period is that public and private organisations can contribute cyber staff with only a minor impact to their own operations.  While this would not guarantee contributions, it would arguably make them more likely.

\begin{figure}[ht]
\centerline{
\includegraphics[width=\columnwidth]{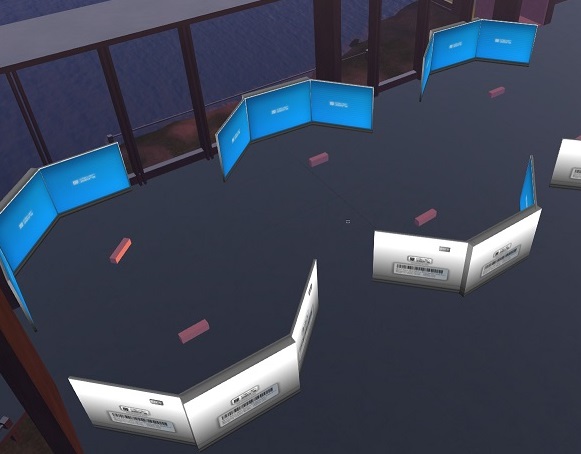}}
\caption{PoC VCE Layout}
\label{fig:PoC1}
\end{figure}

A proof of concept Cyber OMR VCE was created using OpenSimulator, following the design given in figure~\ref{fig:VCE-Layout-Type2}.  Figure~\ref{fig:PoC1} shows the result.  The environment is split into six areas, each one being an area for an AoR team.

\begin{figure}[ht]
\centerline{
\includegraphics[width=\columnwidth]{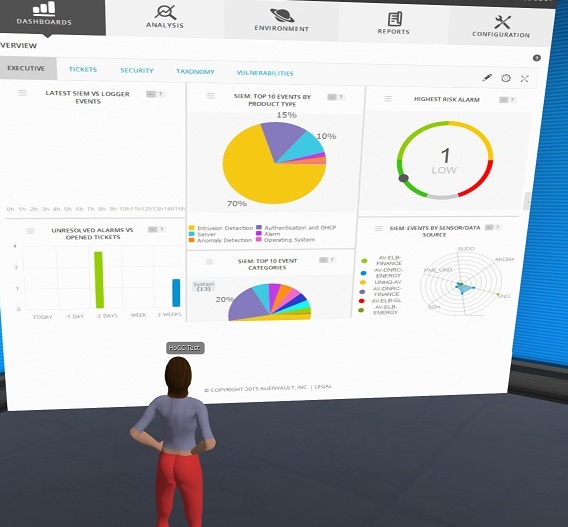}}
\caption{PoC VCE Resource Sharing}
\label{fig:PoC2}
\end{figure}

The VCE allows cyber peacekeepers and local stakeholders to log in and be represented as an avatar. They are able to communicate with each other via voice and text and to bring up data on the three displays for interaction and discussion.  In our example, these screens link to the AoR sensors and display sensor data.  An example of this is shown in figure~\ref{fig:PoC2}.  In practice, these displays can be used for sharing any kind of data to aid in discussion and analysis.  In effect, it is a virtual security operations centre (SOC) but one that can be quickly deployed and used by multiple remote analysts.

Whilst we have proposed that a VCE is used to mitigate the challenge of securing cyber staff contributions, we have noted that it brings other benefits.  These benefits are summarised as follows:

\begin{enumerate}
	\item Encourages cyber staff contributions - Nations will be more likely to contribute cyber expertise if that expertise will not be lost at home.  Contributions of time can be a fixed amount of hours per day.
	\item Transparency - Cyber peacekeepers can be open in their activities by inviting local stakeholders into the environment to witness their work.
	\item Mission cohesion - The VCE brings all cyber peacekeepers in a region together, allowing cross correlation of events and sharing of expertise.  Non cyber units can also be present, to aid mission wide cohesion, communication and unity of effort.
	\item Agility and Cost - The VCE can be brought online with minimal cost.  There is no need to house or feed remote cyber peacekeepers.
	\item Safety - Remote cyber peacekeepers in the VCE cannot be physically harmed.
	\item Training - New cyber peacekeepers can be vetted and trained inside the environment.
\end{enumerate}

Considering all of these advantages, it is concluded that the VCE will bring benefits beyond just OMR. It will therefore become a central tool for conducting multiple aspects of future peace operations.

\subsection{Secure Communications}\label{SecureComms}
In the case of remote cyber OMR, the sensors will be located locally in the conflict area whilst those analysing the data will be located remotely.  A secure means of transferring the data collected by sensors to the central server and subsequently into the VCE is therefore required.  Similarly, in the case of the local approach, locally based cyber peacekeepers will need a secure channel to transmit their reports.

Due to the nature of cyber warfare, we cannot automatically assume that the technology and infrastructure we rely upon to perform this task will be secure.  Hardware, software and public communications networks could potentially be compromised~\cite{josang2014}.  In cases of ongoing cyber attacks, networks could be flooded with traffic and routing systems compromised~\cite{chakrabarti2002}.  It is therefore proposed that a cyber peacekeeping unit will require a robust set of technologies to mitigate these issues.

To begin, a dedicated and mobile communications link which is difficult for third parties in a region to tamper with will be required.  A potential solution here would be the use of satellite services.  For example, the Broadband Global Area Network (BGAN) offered by Immmarsat~\cite{Immarsat2017} can provide cyber peacekeepers with speeds of up to 492kbps.  Coverage is close to global and the terminals are approximately laptop size and weight.  This solution would create a reasonably secure communications path for cyber peacekeepers to operate inside a region and perform their duties, regardless of the condition of local networks.  Research into the security and privacy of mobile networks is an ongoing area of study and developments here can also bring solutions~\cite{Ferrag2018}.  We must also consider the hardware used by cyber peacekeepers; components such as CPUs and motherboards also have the potential to be compromised along the supply chain~\cite{josang2014}.  This is a much harder threat to mitigate, and is a common problem.  For example, in 2010 Dell Power Edge 410 servers were shipped with malware pre-installed on the motherboards~\cite{boyson2014}.  Establishing a secure supply chain will consequently be a challenge faced by a cyber peacekeeping unit.

\subsection{Proof of concept tools}
We have purposely avoided listing specific tools that cyber peacekeepers should use to perform OMR and there are two primary reasons for this.  Firstly, cyber security tools are always evolving, and any list provided here would be quickly out of date by the time cyber peacekeeping is needed.  Secondly, cyber peacekeepers are being employed for their expert knowledge, and should be allowed to select the tools and monitoring methods that suit their specific site.  However, for the purposes of exploring the practicalities of cyber peacekeeping OMR, we briefly experimented with potential tools that could fulfil the cyber OMR goals of our scenario.  

To recap, the value we aim to bring is in detecting actual or impending threats to civilian security.  In the case of the power plant, this could be through a power cut, power surge or a violent event such as an explosion by tampering with the logic that runs the plant.  Achieving this goal will be supported by monitoring for changes in network structure and traffic (i.e. raising situational awareness in an AoR).

In our proof of concept environment, we already have sensors collecting data and a VCE to view and analyse the results.  We considered ways to monitor the network structure, and found that commonly available tools such as Alienvault OSSIM had functionality to monitor the availability of hosts.  When a particular host was taken down, alerts could be raised and viewed in the VCE.  Important structural devices such as routers and firewalls could be configured to send their logs to the SIEM and custom rules developed to monitor for changes.  Hence, any changes to the network structure could be highlighted in the VCE, allowing analysis of the change.  Other tools such as the Passive Real-time Asset Detection System (PRADS) can also be used to monitor for new devices or changes in known asset behaviour~\cite{Sanders2013}. 

\begin{figure}[ht]
\centerline{
\includegraphics[width=\columnwidth]{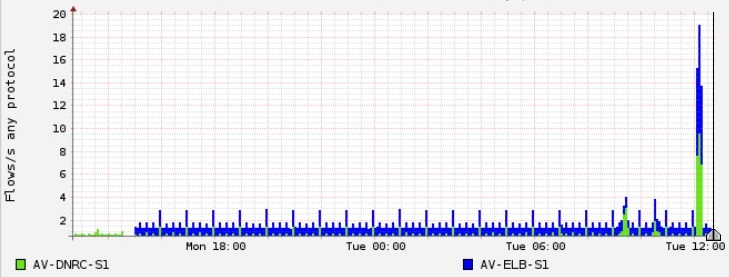}}
\caption{Detecting a change in network traffic volume}
\label{fig:netflow}
\end{figure}

Similarly for changes in network traffic, we were able to configure NetFlow~\cite{Cisco2012} on the sensors.  This allowed analysis of the network traffic volume over time.  An example of a simulated change in network traffic is shown in figure~\ref{fig:netflow}.

While it is beyond the scope of this article to compare and test specific monitoring tools, we have shown that there are many practical tools to perform the technical aspects of cyber OMR in an ICT environment if it was required today.  Where more of a challenge will be found is in operational technology (OT) environments such as critical national infrastructure.

\section{Cyber OMR at Critical National Infrastructure}
The methods and techniques described thus far will be effective in a traditional ICT environment such as the president's office scenario.  Where monitoring becomes more challenging is in OT environments such as critical national infrastructure (CNI).  Power grids, public water supplies and transport networks will be commonly requested sites for cyber OMR.  This is because attacks upon these systems present a significant threat towards life and to the ongoing stability of a nation.  In this section, we therefore describe the characteristics of CNI from a monitoring perspective, the challenges that will be faced and ways to tackle those challenges.

\subsection{CNI Background}
The UK government defines national infrastructure as the "facilities, systems, sites, information, people, networks and processes, necessary for a country to function and upon which daily life depends.  It also includes some functions, sites and organisations which are not critical to the maintenance of essential services, but which need protection due to the potential danger to the public (civil nuclear and chemical sites for example)."~\cite{CPNI2018}.  Figure~\ref{fig:CNIareas} shows which sectors are identified by the UK as national infrastructure:

\begin{figure}[ht]
\centerline{
\includegraphics[width=\columnwidth]{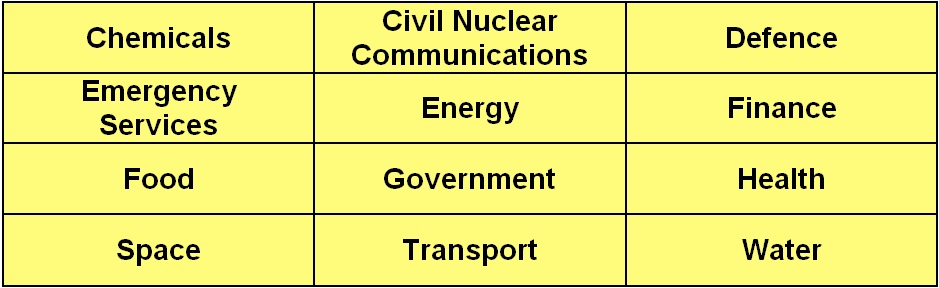}}
\caption{Areas the UK regards as National Infrastructure}
\label{fig:CNIareas}
\end{figure}

Critical National Infrastructure is defined as:  "Those critical elements of national infrastructure (facilities, systems, sites, property, information, people, networks and processes), the loss or compromise of which would result in major detrimental impact on the availability, delivery or integrity of essential services, leading to severe economic or social consequences or to loss of life."~\cite{CPNI2018}.  This definition supports our assertion that CNI will be a popularly requested site for cyber peacekeeping, since failure directly impacts peace and security.

Research into the cyber security of CNI and the challenges it presents is extensive~\cite{Robinson2013,cornish2012,rudner2013,cook2016}. Piggin~\cite{Piggin2014} provides a concise overview of the potential impacts and attack vectors, but it is prudent to give a brief history and overview of the main challenges, to enable discussion on how to conduct cyber OMR at such sites.

National infrastructure of the past was generally isolated from the outside world, operating as standalone entities with no external connections~\cite{Robinson2013}.  In this isolated environment, the threat of an external actor performing some kind of malicious act was small.  Hence, the availability of the systems was the top priority, with confidentiality and integrity being less important.  The protocols used between devices at such sites (Modbus, DNP3 etc.) reflected this:  encryption was rare and communications were not tamper resistant~\cite{alcaraz2015}.

With the arrival of affordable computing and networking, owners of these facilities saw benefits in connecting them together.  Rather than employ staff to monitor a single facility, geographically distributed plants could be monitored and operated from a central location using Supervisory Control and Data Acquisition (SCADA) systems~\cite{Robinson2013}.  This provided a greater level of control with improved efficiency.  Today, national infrastructure capitalises on many modern technologies including mobile networks, wireless communications, the internet and embedded devices~\cite{alcaraz2015,Sadeghi2015}.  Whilst the technology used at CNI can be referred to by a number of terms, we adopt the umbrella term of operational technology (OT).

Whilst the OT used in CNI has enjoyed improved connectivity and efficiency, it has effectively been wrapped around legacy protocols and devices that have remained unchanged from the days of isolation.  This has created the perfect storm of introducing multiple vectors of attack into systems which generally have poor security features.

\subsection{Challenges of cyber OMR at CNI}
We have argued that CNI will be a significant part of cyber OMR for two reasons.  Firstly a failure of CNI has the potential to seriously damage peace and security in a region through harmful impacts~\cite{Piggin2014}.  Secondly, it is particularly vulnerable to cyber attack due to increased interconnectedness and weak security at core components.  We must therefore explore how cyber OMR could effectively monitor such sites.  From the perspective of conducting OMR, we propose that cyber peacekeepers will face the following CNI specific challenges:

\begin{enumerate}
	\item Fragility - older OT hardware can be fragile~\cite{merabti2011}.  Software used to program ladder logic can be basic with poor error handling.  A network port receiving unexpected data can cause a hardware reset, in some cases causing a loss of the device's current logic.  Research has shown that tasks such as asset discovery can only safely be performed via passive methods~\cite{Wedgbury2015}.

	\item Bespoke attacks - Sensors on traditional ICT networks use signature based detection to identify known attacks.  While this technique will detect some attacks against OT systems (particularly machines running common software such as Microsoft Windows or Java), it will not help in cases where there is a targeted attack against specific OT hardware.  Stuxnet was a bespoke cyber weapon crafted specifically for attacking Iranian nuclear facilities~\cite{kushner13}, for which no signature existed.
	
	\item Time critical operation - Some OT systems are time critical in their operation.  A certain action must occur at a certain time, with any delay having knock on effects to later processes. Delays of milliseconds before a message is delivered has the potential to cause problems.
	
	\item Downtime unacceptable - In ICT environments downtime for maintenance is acceptable.  It can occur at times where it will cause the least inconvenience.  In an OT environment, this is not true.  The power grid cannot simply be taken down for five minutes, it is an essential service that is required 24/7.
	
	\item Proprietary protocols - Observing TCP/IP networks is well understood: the structure of messages is known and documentation is freely available.  Protocols used by older OT devices are often proprietary, for which limited or no documentation is made available by the manufacturer.  Some of these protocols are no longer supported but are still used, due to the hardware remaining in service for decades.  This can not only lead to no support from the vendor, but also a declining pool of expertise as staff who are familiar with the protocol retire or change jobs.
	
	\item Airgapping - In the past, manufacturers of OT equipment recommended that their hardware should be airgapped: having no physical connection to the outside world.  While a good idea in theory, airgapping is now considered impractical~\cite{Byres2013}.  However, CNI owners may still abide by this concept and refuse any outside connections to the site.
\end{enumerate}

To explore the feasibility of performing cyber OMR at CNI, we conducted practical exercises to determine what is possible with existing tools.

\subsection{Exploring the practicalities}
Performing research upon OT systems is challenging. As discussed, they are often fragile and have a zero-downtime requirement.  CNI owners are understandably reluctant to allow researchers access to conduct experiments at critical sites.  A common solution is to use simulations: using software to represent the hardware, software and protocols that would be in use at a real site.  Simulations of OT systems have been developed by other researchers~\cite{Queiroz2011,Mallouhi2011}, but are limited in their ability to reflect all of the properties of an operational deployment~\cite{Hahn2010}.  An alternative approach is to develop a test bed using the actual hardware and software found in critical national infrastructure.  This approach brings benefits: a simulation can represent what hardware is supposed to do, actual hardware shows both documented and undocumented behaviour.  The primary disadvantage in this approach is cost.

\begin{figure}[ht]
\centerline{
\includegraphics[width=\columnwidth]{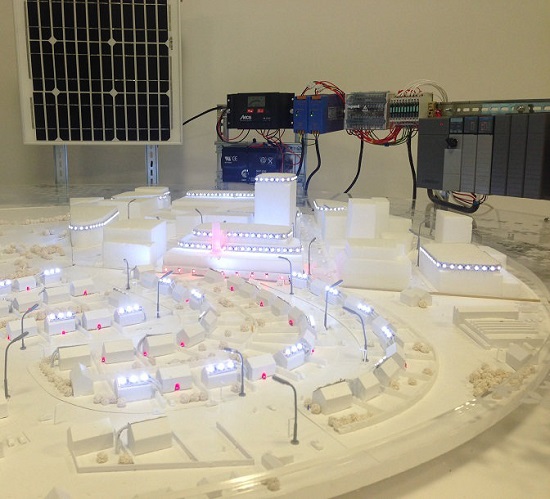}}
\caption{Airbus OT test bed city}
\label{fig:city}
\end{figure}

The OT test bed based at Airbus in Newport (Wales) is such a solution, and was made available to assist in our research.  As part of Airbus' wider research into OT security, the testbed is used to explore new ways of protecting the core devices which run CNI and the team have a number of OT specific attacks to test novel defences.  This made it a good choice for also testing how cyber peacekeepers could practically monitor for such attacks.   Linking back to our scenario, the testbed's smart city was used to represent Country A's power grid.  This is a model of a city, where the power supply is controlled by Allen Bradley controllers and the Ignition HMI software.  The city is shown in figure~\ref{fig:city}.

To test traditional tools, an ICT based sensor was built (based upon Alienvault OSSIM) and added to the test bed.  A free port was located on the network, and a mirror configured to forward traffic.  While this was easy to accomplish in a test bed environment, it will be more complex in a live environment.  Site operators may be suspicious of adding new devices to what may be a fragile system that has remained unchanged for years.  Finding a physical connection into the network may be challenging, and connecting the sensor into a position where it can see traffic without interrupting or delaying communication will be an issue.

Once the sensor was in place, an attack was launched against the control system resulting in a loss of power to the city.  The sensor did not raise any alerts during this attack.  This was the expected result, since modules were designed to detect threats in ICT, not OT, environments.  Other researchers have found similar results~\cite{Mahboob2013}.

Commercial security product vendors have noticed this gap in the market, and a number of products are starting to enter the market for OT monitoring.  For example, Check Point Software Technologies\cite{Checkpoint2016}, Alienvault~\cite{Alienvault2018} and Claroty~\cite{Claroty2018} advertise specialised OT security monitoring tools.  Researchers have recently developed distributed intrusion detection system (DIDS) for SCADA systems~\cite{cruz2016cybersecurity}. Airbus itself has also developed a number of prototype solutions to address specific OT security issues, such as safe asset discovery and forensic investigation tools.

Looking wider, further possibilities include model-based detection~\cite{Cheung2007}.  Here the behaviour of an OT system is passively monitored for abnormalities in its operation.  Researchers such as Nicholson, Janicke and Cau~\cite{Nicholson2014} have explored this, by examining the value of Interval Temporal Logic (ITL) as a method for observing the state of OT hardware and software systems at various points in time.  They launched two attacks against a controller, and noted that ITL was successful in spotting that the state of the controller had changed.  XSense from CyberX claims to leverage this approach using machine learning, modelling abilities and a patented state machine design~\cite{CyberX2017}.  The product requires a learning period, where normal operation is witnessed and recorded by the tool.  In a cyber OMR context this "model based" approach may have limited utility; the ability to provide a period of learning in a "safe" state may be lacking if the site is already compromised before cyber peacekeepers arrive.

Any attempt to define specific tools or methods to conduct cyber OMR at CNI would be flawed: effective monitoring will depend upon the configuration of hardware, protocols and software at a specific site.  Furthermore the quality and quantity of OT monitoring tools is advancing rapidly in this area and any recommendations would quickly become out of date.  We therefore propose that cyber OMR at CNI will require a bespoke monitoring solution, where the expertise and knowledge of cyber peacekeepers is applied along with OT vendor collaboration.  As an example, there are specific methods for monitoring power distribution sites such as the deployment of phasor measurement units (PMUs) to measure the actual voltage and other variables in real time~\cite{eto2015,jamei2016}.  Such a solution was a major recommendation following the 2003 blackout in the northeastern United States~\cite{liscouski2004} and will require specialist knowledge to implement.

\section{Reporting}\label{Reporting}
Previous sections have described how to perform cyber observation and monitoring.  The aim of this section is to explore the practicalities of the final aspect: reporting.  Reporting is just as important as the monitoring and observation that comes before it.  We can gather all kinds of valuable observations, but that value is lost if they are not communicated properly and used in decision making.

\begin{figure}[ht]
\centering
\includegraphics[width=\columnwidth]{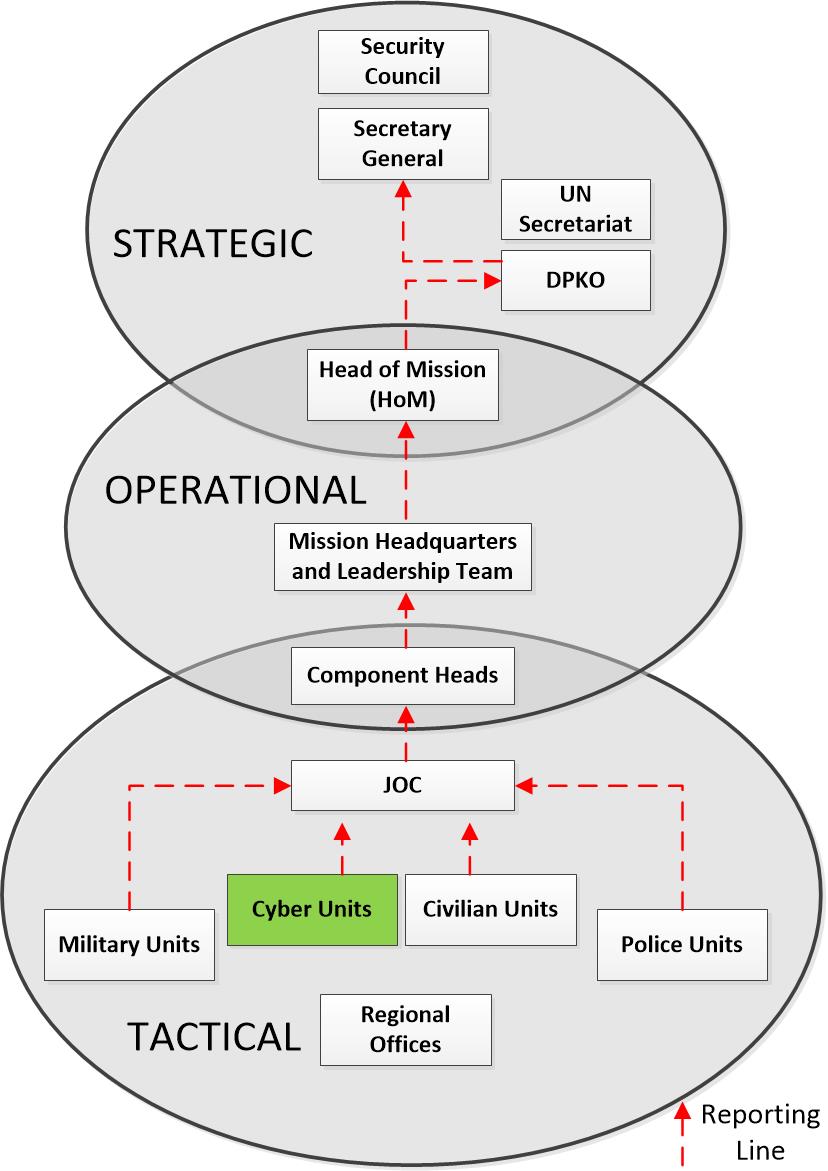}
\caption{UN Operational Reporting System with Cyber added}
\label{fig:reportingsystem}
\end{figure}

UN peacekeeping doctrine states that reports should be "timely, accurate, clear and concise, substantiated with evaluations and assessments, to support higher commanders' decision-making"~\cite{UNIBAM2}.  UN standard operating procedures for reporting are well documented~\cite{UNPDT2010}, and as with all aspects of cyber peacekeeping, we aim to fit cyber into this existing process.  In a UN peacekeeping operation reports flow up from the tactical level up to the strategic level, with various points in between for filtering and collation of data.  Using our modified organisational structure from section~\ref{organisation}, figure~\ref{fig:reportingsystem} shows how cyber fits into this existing system. 

Peacekeepers in a cyber unit will submit reports to the Joint Operations Centre (JOC), which already collates and cross-references information coming in from all units.  This allows the JOC to formulate a daily report, providing the component heads and leadership team with a coherent overview of the day's events at the tactical level.

In general, there are two types of reports that cyber peacekeepers performing OMR will need to submit:  Daily SITREPS and Special Incident Reports.  Cyber peacekeepers will be observing their AoR in order to detect and observe for a variety of events.  Some of these events will be non-critical, such as a decline in cooperation from local staff or a non-malicious and non-critical repeat failure of a particular device at CNI.  These events can be reported in the daily SITREP.  This is a situation report which is submitted to the JOC on a daily basis, regardless of whether the peacekeeper considers there to have been any reportable events.  The process of creating and forwarding daily SITREPS is well established in peacekeeping training material~\cite{UNPDT2010} and do not require any alterations for cyber.  At the end of the day, cyber analysts based at the JOC meet with their team and convey their summary of cyber events for inclusion into the overall daily situation report sent to the operational level.  Cyber peacekeepers are also observing for events which could lead to a threat to peace (e.g. to civilian life).  In such cases it will be necessary for the event to be reported quickly.  Here a special incident report (SINCREP) will be necessary, which is immediately submitted to the JOC for attention.

\subsection{SINCREP Example}
To provide a concrete example of our proposed system, we return to our scenario.  Let us assume that the power plant AoR team detects unusual network activity on the power plant's control network.  Abnormal messages are being sent to logic controllers, but peacekeepers at this stage are not sure what the effect will be.  They generate a SINCREP which communicates their observations.  They include the time the unusual activity first started, that it is ongoing, the IP addresses and hardware involved, a description of what is happening and the network capture files as evidence.  This report is sent to the JOC and received by cyber staff.  They analyse the report and view the evidence, entering into the virtual environment to communicate with the reporting peacekeepers in real-time.  Because the JOC is collecting reports from all components, the cyber staff notice that a SINCREP is received from a police unit that power was briefly lost in a certain region.  Putting these reports together to form a picture of the whole situation, the JOC is able to provide advance warning to other components that blackouts might be imminent around the region.  The JOC then acts as a crisis centre to monitor the situation, inform relevant stakeholders, whilst the cyber peacekeepers continue to monitor the event and provide advice to local staff on defensive measures to take.

This scenario brings to the fore a significant question:  should peacekeepers performing cyber OMR stand by and simply monitor and report whilst a situation degrades and threatens civilians, or should they actively intervene?  This is not a new question faced by peacekeeping, and has undergone much discussion and debate~\cite{Ryngaert2015,wills2009,UN-A68767}.  Many past criticisms of UN peacekeeping have touched upon either the inability or unwillingness of peacekeepers to use force to protect civilians around them.   UN Resolution 1265 (1999) emphasised the importance of protecting civilians during peace operations.  Many missions have subsequently been established with a specific mandate to protect civilians from harm, using robust measures and loosening up use of force controls.  This has led to the UN moving away from traditional doctrine, the Brahimi report and core principles in favour of more robustly protecting civilians~\cite{deConing2017,ramos2015}.

With this shift of focus in mind, we will not attempt to state that a cyber peacekeeper performing OMR at a critical site should passively observe as a threat to civilian security materialises.  Such an act would arguably be a repeat of past failures such as Bosnia~\cite{BBC2014} or Rwanda~\cite{barnett1999}.  It is therefore accepted that when faced with an event where there is potential for harm to civilians, cyber peacekeepers should take any reasonable steps to prevent that harm if they are able to.  The impacts of this decision will be further considered in the evaluation.

\subsection{Integration with the VCE}
Since we have proposed that a virtual environment will be central to cyber peacekeeping, it is prudent to explore the possibility of integrating the reporting system into it.  The goal is to have as many tools as possible to perform cyber OMR in one unified location.

As a proof of concept, we used the RequestTracker~\cite{RQ2016} ticket system.  This software allows immediate transmission and handling of reports, fulfilling the requirement that reports are timely.  Custom fields were added to the ticket interface, with the aim of ensuring that tickets fulfilled the requirement for reports to be concise.  The following fields were added:

\begin{itemize}
  \item Start time of event.
	\item Type of event.
	\item Supporting evidence.
	\item AoR the event was detected in.
	\item The sensor which detected the event.
\end{itemize}

\begin{figure}[ht]
\centering
\includegraphics[width=\columnwidth]{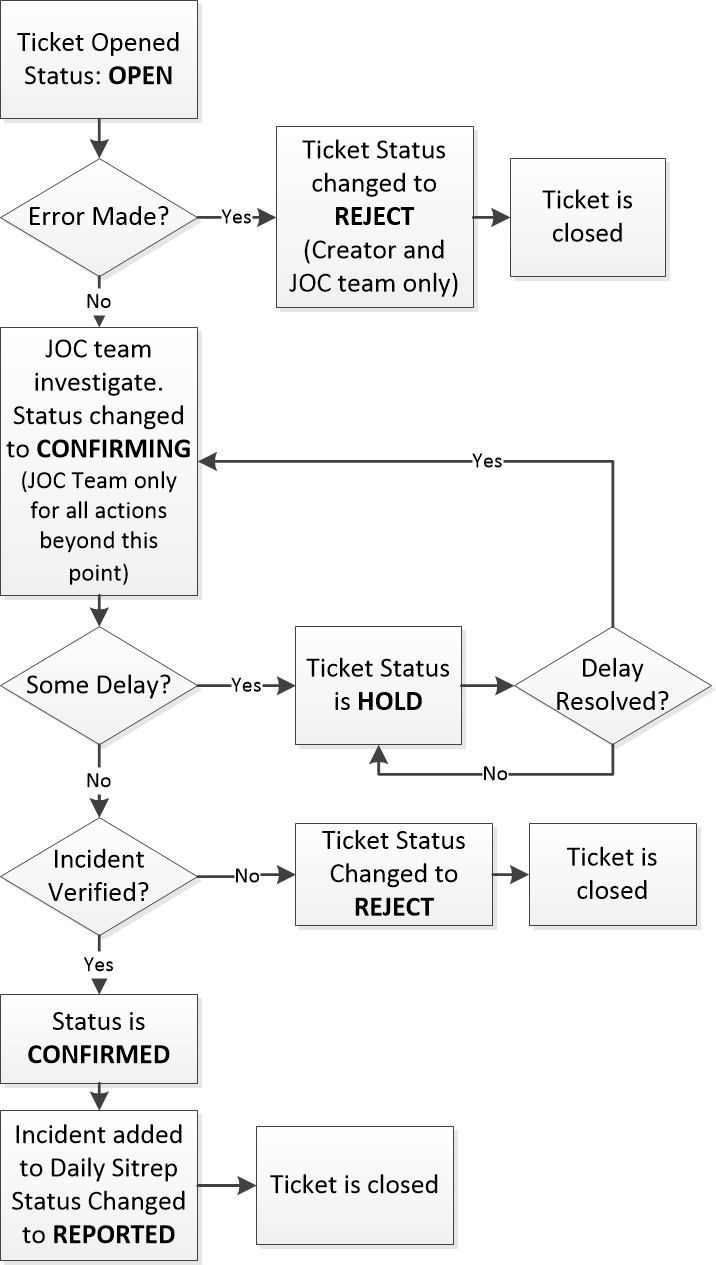}
\caption{Proposed Cyber OMR Ticket Lifecycle}
\label{fig:TicketFlow}
\end{figure}

Some further modifications were made, with the end result being a proof of concept cyber OMR reporting system.  The next goal was to examine how the ticket is managed post-creation.  To define a set of rules regarding ticket states for cyber OMR, it was first necessary to develop the life cycle of a cyber peacekeeping OMR ticket.  The proposed life cycle is presented visually in figure~\ref{fig:TicketFlow}.

\begin{figure}[ht]
\centering
\includegraphics[width=\columnwidth]{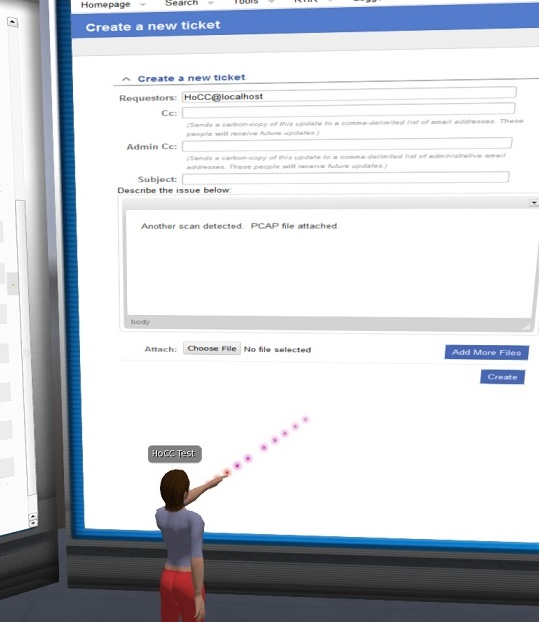}
\caption{Interacting with RT from within the VCE}
\label{fig:RTinVCE}
\end{figure}

It was found that the ticket system integrated well into the VCE.  Reports could be written inside the environment, allowing multiple actors to contribute to the content.  An example is shown in figure~\ref{fig:RTinVCE}.

\section{Evaluation of Cyber OMR against UN Peacekeeping Principles}
In this paper, we have explored practical ways in which cyber OMR could be performed.  We now evaluate our proposals against the core UN peacekeeping principles of consent, impartiality and non-use of force except in self defence or defence of the mandate.

\subsection{Consent}
UN peacekeeping doctrine~\cite{UN2008} states that consent of the parties is one of the core UN principles.  It is thus necessary to ensure that cyber OMR as we have designed it does not have the potential to violate this principle.

In our idealistic scenario, the presence of consent is clear.  Both parties have requested UN assistance in overseeing the cyber warfare aspect of the peace agreement, and both sides have invited the UN to monitor certain critical sites.  Consent in this case is not only respected, but we would argue fundamentally required.  Attempting to install sensors, gain familiarity with sites and keep the system running is only possible by working closely with local staff.  If local staff are not cooperative or work against cyber peacekeepers, cyber peacekeeping OMR will be extremely challenging to perform effectively.  Consent is the reason why some observational goals were deemed to be low feasibility. For example, monitoring for violations of privacy.  If a host nation is snooping upon its citizens in the cyber domain, they are unlikely to consent to cyber peacekeeping if it was believed that cyber peacekeepers were looking to criticise privacy violations.  Therefore, the principle of consent leads to such activities not being included at this stage.  It is not worth blocking the value that cyber OMR can bring in regards to protecting civilian lives, for the sake of attempting to also protect their privacy.

It is foreseeable that there will be cases where the principle of consent can become threatened.  For example, peacekeepers performing cyber OMR at sites such as nuclear plants or dams could face a situation where a significant threat to civilian life has been detected but local staff refuse and resist corrective action to prevent it.  This will be a problem, since it is made clear in UN peacekeeping doctrine that peacekeepers must not stand by and remain passive in the face of clear threats to civilians~\cite{Brahimi2003}.  We have already proposed that a cyber peacekeeper should not stand by and passively observe while an event escalates to a level where civilian life is threatened.

In such cases, there is strong justification for intervention without consent: effectively a step into peace enforcement.  How this could be done remains an open question, and it could have significant negative effects such as a withdrawal of consent for the whole peace operation.  Planning could minimise the likelihood of this risk.  For example, parties requesting cyber OMR must agree at the outset that threats to civilian life must be acted upon by local staff, and that interventions may take place if cooperation at these times is lacking.  By making parties aware of this, the effects of these interventions can be minimised to not endanger the whole peace process.  Drills of a crisis situation could be valuable, to gauge how local staff react and highlight any potential issues before they arise.

A further complication to consent is the area of public versus private ownership.  While kinetic OMR primarily takes place in the public domain (observing roads, bridges, towns etc.), cyber OMR has the potential to be taking place in privately owned networks.  For example, a nuclear power plant may be owned by a foreign energy company.  This means that whilst kinetic OMR can function purely with the consent of a national government (the owners of these public spaces), cyber OMR will potentially require the consent of both public and private entities.  This is an interesting area to consider and must form a suggestion for future work.

Finally, whilst our scenario represents an ideal scenario from the perspective of gaining consent, the reality of peacekeeping is often less clear.  As noted by de Coning~\cite{deConing2017} peace operations today have been noted for their lack of clear consent, with the Security Council issuing a mandate against a specific party to the conflict.  If our scenario changed so that Country B did not provide consent, the task of performing cyber OMR in that country would become much more challenging.

\subsection{Impartiality}
The principle of impartiality is important to UN peacekeeping, ensuring that peacekeepers can act as a trusted party.  In our scenario, impartiality is respected.  Both sides are offered cyber OMR, and peacekeepers are not attempting to attribute cyber attacks to a particular side.

Cyber OMR must be offered to all parties of a conflict equally.  If it appears that the UN is providing cyber OMR for a number of sites in Country A and that Country B is not offered the same opportunity, there is potential for the principle of impartiality to be violated.  It should therefore be made clear to parties at the outset that cyber OMR is available to all, but that only a limited number of sites will be monitored based upon the potential threat to civilian security.

A threat to impartiality is found in resource contention.  One party may request cyber OMR at twenty sites, consuming the majority of UN cyber peacekeeping resources.  If another party makes a similar request a month later, the UN may find itself unable to fulfil the request and impartiality will again be violated.  Peacekeepers must be conservative when agreeing to monitoring, and only agree to sites where failure could lead to civilians being harmed or another threat to peace, such as state collapse.  The focus of cyber OMR must clearly be one of these goals, not to provide "free" security monitoring so that a party can divert their own cyber resources elsewhere.

Cyber OMR teams must also make efforts to be transparent in the tasks they perform, to avoid claims that cyber peacekeepers are helping one party more than the other.  Considering our proposal that cyber peacekeepers must intervene if they can prevent a threat to civilians, this will be challenging.  Our virtual collaborative environment presents opportunities here, with the potential for stakeholders from both sides to have a presence in the environment and witness the tasks that cyber peacekeepers perform.  This would enhance transparency, and help to minimise claims of partiality.

\subsection{Non use of force}
Our design of cyber OMR does not utilise force.  Attacks against an AoR are observed and reported through the reporting system.  Cyber peacekeepers are not attempting to take offensive actions such as launching counter cyber attacks.  Although OMR is primarily passive, it was noted that peacekeepers are required to intervene if a grave violation of human rights is observed.  This equates to a situation such as a cyber attack which is close to opening a dam or making the public water supply toxic.  In such scenarios, cyber peacekeepers must intervene if possible to prevent the harm.  This intervention will likely be through notifying the network owner, advising them on what must be done and assisting with implementing the action.  Such an action would not be a use of force.  In cases where the response from the network owner is lacking, local cyber peacekeepers may have to take enforcement action.  If the local staff actively resist, there is a situation where force could be the next step.  This would require cyber units to be supported by police or military units to provide physical security whilst the action is taken.  In such a scenario, the use of force would be in defence of the mandate.  It would therefore not violate the principle.

\section{Conclusions and Future Work}
In this paper, we have explored the practicalities of starting up and performing just one cyber peacekeeping activity:  cyber OMR.  Basing our work on the foundations set in previous work~\cite{Robinson2018} we have reached a number of conclusions, summarised below:

\begin{itemize}
    \item Cyber terms in peace agreements should be chosen carefully.  We want to avoid terms which require solid attribution in cyberspace, and favour those which can be measured at a human interaction level while still bringing value.
    \item Securing the required cyber expertise will likely be the biggest obstacle towards cyber peacekeeping.
    \item Cyber fits easily into existing structures and processes.
    \item Cyber OMR will bring most value at CNI, with a focus on protection of civilians and state stability.
    \item Technical obstacles towards monitoring CNI are being broken down as new products and tools come to market, but there is still a skills shortage in this area, which will place further pressure on securing capable staff.
    \item Use of a virtual collaborative environment brings a number of benefits including transparency, ease of collaboration, information sharing and the potential for states to contribute their cyber experts without losing capability at home.
    \item Our proposals do not violate established UN peacekeeping principles.
\end{itemize}

The discussions held in this paper raise deeper questions about cyber OMR as an activity.  Firstly, UN peacekeeping as a whole is currently undergoing a shift in how it operates.  The doctrine we have based our scenario on is the 2008 capstone~\cite{UN2008}.  This is the current 'official' way in which UN peacekeeping should work, based upon the findings of the Brahami report~\cite{Brahimi2003}.  However, many authors point out that UN peace operations of today do not follow this doctrine~\cite{deConing2017,Mateja2015}. For example, in the DRC consent from all parties is not present, with some openly hostile to peacekeeping forces.  Neither is impartiality, with the UN effectively supporting a government against insurgency.  Non use of force is also questionable, with the UN actively partaking in offensive operations in collaboration with government forces~\cite{deConing2017}.  It would therefore be useful to present a scenario which does not fit the 2008 doctrine and explore the impact upon the value and feasibility of cyber OMR.

We proposed that a peacekeeper performing cyber OMR must intervene to prevent harm if they are able to do so.  This raises a question about the role of cyber OMR as a passive activity.  If cyber peacekeepers are commonly stepping in to take action, it is arguably more efficient to establish a cyber buffer zone from the outset.  In this regard, it is possible to propose that cyber OMR should not be a standalone activity, but rather a component of a cyber buffer zone.  This will be explored in future work regarding cyber buffer zones.

An area of future work is to look at feasible ways of solving the challenge of securing cyber expertise.  If the major obstacle towards effective cyber OMR will be the limited supply of cyber expertise and political concerns from highly developed nations (as discussed in section~\ref{findingexpertise}), it would be prudent to explore non-UN means of protecting civilians and state stability both during and following cyber warfare.  For example, alliances such as NATO or the Arab League where the member states share common military goals and would be more willing to share cyber expertise.

It would also be valuable to explore how existing frameworks could be leveraged to bolster cyber OMR.  For example, the NIS Directive in Europe demands that member states have in place a framework and national cyber security authority (NCSA) so that they are equipped to manage cyber security incidents~\cite{Maglaras2018,piggin2018}.  These NCSAs would likely be major enablers and contributors to cyber peacekeeping, allowing peacekeepers to quickly enter sites and gain familiarity.  Operators of essential services also have to take appropriate security measures and are required to report cyber security incidents.  In this regard, it could be argued that frameworks such as the NIS Directive already go some way to motivating owners of critical infrastructure to develop a monitoring capability.  From a cyber peacekeeping perspective, efforts such as NIS could reduce the initial workload to establish a monitoring capability in a nation.

Finally, this article has only sought to develop the peacekeeping activity of cyber OMR.  As described in Robinson et al.~\cite{Robinson2018}, there are many more valuble and feasible activities which need further development.  Developing activities such as DDR, SSR, malware action and the practicalities of cyber ceasefire agreements remains an open area of work.  Input from experts in these areas is essential, since cyber peacekeeping is fundamentally a cross disciplinary challenge.  Furthermore, other forms of peace operations such as conflict prevention and peace enforcement also remain open.  Conflict prevention in particular has been regarded as critical towards the future of peace operations~\cite{Annan2001}.  It would therefore be valuable to explore efforts such as the confidence building measures put forward by the OSCE~\cite{OSCE2016} and other groups.

\ifCLASSOPTIONcaptionsoff
  \newpage
\fi
\balance
\bibliographystyle{ieeetr}
\bibliography{refs}

\end{document}